\documentclass[aps,prl,twocolumn,superscriptaddress]{revtex4}

\usepackage{graphicx}
\usepackage{dcolumn}
\usepackage{bm}

\begin{document}


\title{Variations in Substitution Rate in Human and Mouse Genomes}


\author{H.H. von Gr\"unberg}
\affiliation{Universit\"at Konstanz, Fachbereich Physik, 78457 Konstanz, Germany}
\author{M. Kollmann}
\thanks{Corresponding autor: kollmann@lagash.dft.unipa.it}
\affiliation{Universit\`a di Palermo, Dip. di Fisica e Tech. Relative, Palermo, Italy}


\date{\today}

\begin{abstract}
We present a method to quantify spatial fluctuations of the
substitution rate on different length scales throughout genomes of
eukaryotes. The fluctuations on large length scales are found to be
predominantly a consequence of a coarse-graining effect of
fluctuations on shorter length scales. This is verified for both the
mouse and the human genome. We also found that both species show similar standard deviation of fluctuations
even though their mean substitution rate differs by a factor of
two. Our method furthermore allows to determine time-resolved
substitution rate maps from which we can compute auto-correlation functions in order to quantify how fast the spatial fluctuations in substitution rate change in time.  
\end{abstract}

\pacs{}

\maketitle


The detailed knowledge of the mechanisms of mutations in living
organisms is of fundamental importance for understanding the evolution
of genomes. On the basis of development of every organism is the cell
reproduction cycle and mutations can be seen as errors made in DNA
during the process of chromosome replication. Mutations occurring in
germ-line cells are inherited and passed onto the next generation. A
large fraction of these mutations are substitutions of single
nucleotides (G,C,A or T) by another while other mutations are due to
insertions or deletions of one or more nucleotides in the DNA
sequence. In recent time there is growing acceptance that the
substitution rate is not spatially constant inside the genomes of
mammals
\cite{wolfe,casane,matassi,nachman,williams,chen,lercher,mouse}. This
is a surprising result at first sight, as nucleotide substitutions
resulting from copying errors should be fairly independent on the
actual position, at least on large length scales. Unfortunately, this
picture is too simple, as there exist strong correlations in the mammalian genome between mutation rate and other evolutionary rates (e.g. recombination rate) \cite{adam,hardison,mouse}. Although the origin of the
substitution rate bias in genomes of mammals is unknown, it is highly important to quantify
the length and time scales where changes are occurring. This is
because the amount of conserved sequences between the genomes of
different species depends crucially on the fluctuations of the local
mutation rate. But these conserved sequences give the best insight how
much of the sequence in genomes has function and is therefore under
selective constraint.

\indent Here, we present a method to calculate substitution rates in genomes of eukaryotes. On grounds of our
method we make use of interspersed repeats \cite{jurka,human,mouse}.
Interspersed repeats are sequences of $3\cdot 10^{2}-5\cdot 10^{3}$
base pairs in length whose copy was inserted up to $10^{6}$ times in
the human genome. Each copying machinery has only worked for a short
time in the history of the organism and from that time on, the copies
of a specific repeat type have accumulated substitutions and other
mutations. Due to the large number of copies the original sequence can
be reconstructed quite accurately in most cases and differences
between a given copy (repetitive element) of an interspersed repeat and its consensus
sequence allows to estimate the mutation rate at the position of this repetitive element if the time is known
when the coping machinery was active (c.f. Revs \cite{jurka}). In the
human genome there have been classified more than $300$ different
interspersed repeats which occupy in total more than $40\%$ of the
genome, which gives us a large set of sequences at hand which is most
likely under no selective constraint.
 We visualize interspersed repeats as ''measurement devices'' for the
underlying local substitution rate. This requires to solve three major
problems: (i) single repetitive elements are usually too short and
have on average accumulated too few substitutions to give a reliable
estimate for the substitution rate at their position in the genome,
(ii) the repetitive elements show a very broad length distribution
(implying that our ''measurement devices'' differ in their
''sensitivity'', which is proportional to their length), and (iii) the
repetitive elements belonging to different types differ in general in
their age (thus, the measurement devices have been measuring over
different periods of time). Now the major theoretical task is to
derive from the large amount of repetitive elements, which have large
diversity in their age and sensitivity, reliable information about the
underlying substitution rate at different positions and at different times in the genome.\\ 
\indent For our analysis we take a set of $M$ different types of interspersed
repeats ($M=200 - 300$). Then, for investigating variations in the
substitution rate on different length scales we divide each genome
(mouse and human) into $Z$ equally sized partitions ($\gamma = 1,
...., Z$). The total number of bases of all repeats of type $\alpha$
in the partition $\gamma$ is denoted by $N_{\alpha\gamma}$ and the
corresponding total number of base mismatches to the consensus sequence originating from single nucleotide substitutions
is given by $k_{\alpha\gamma}$.  One may consider the quantity
$N_{\alpha\gamma}$ as a measure for the sensitivity of the measuring
device, which are repetitive elements of the same type $\alpha$
located in partition $\gamma$. Next, we define for each partition
($\gamma$) and each repeat type ($\alpha$) the average divergence,
$Q_{\alpha\gamma}$, which is the average probability to observe a base
mismatch to the consensus sequence. 
For statistically independent substitutions, the probability to find
$k_{\alpha\gamma}$ mismatches of bases in interspersed repeats of type
$\alpha$ located in the partition $\gamma$, is given by
\begin{figure}
\includegraphics[width=0.45\textwidth]{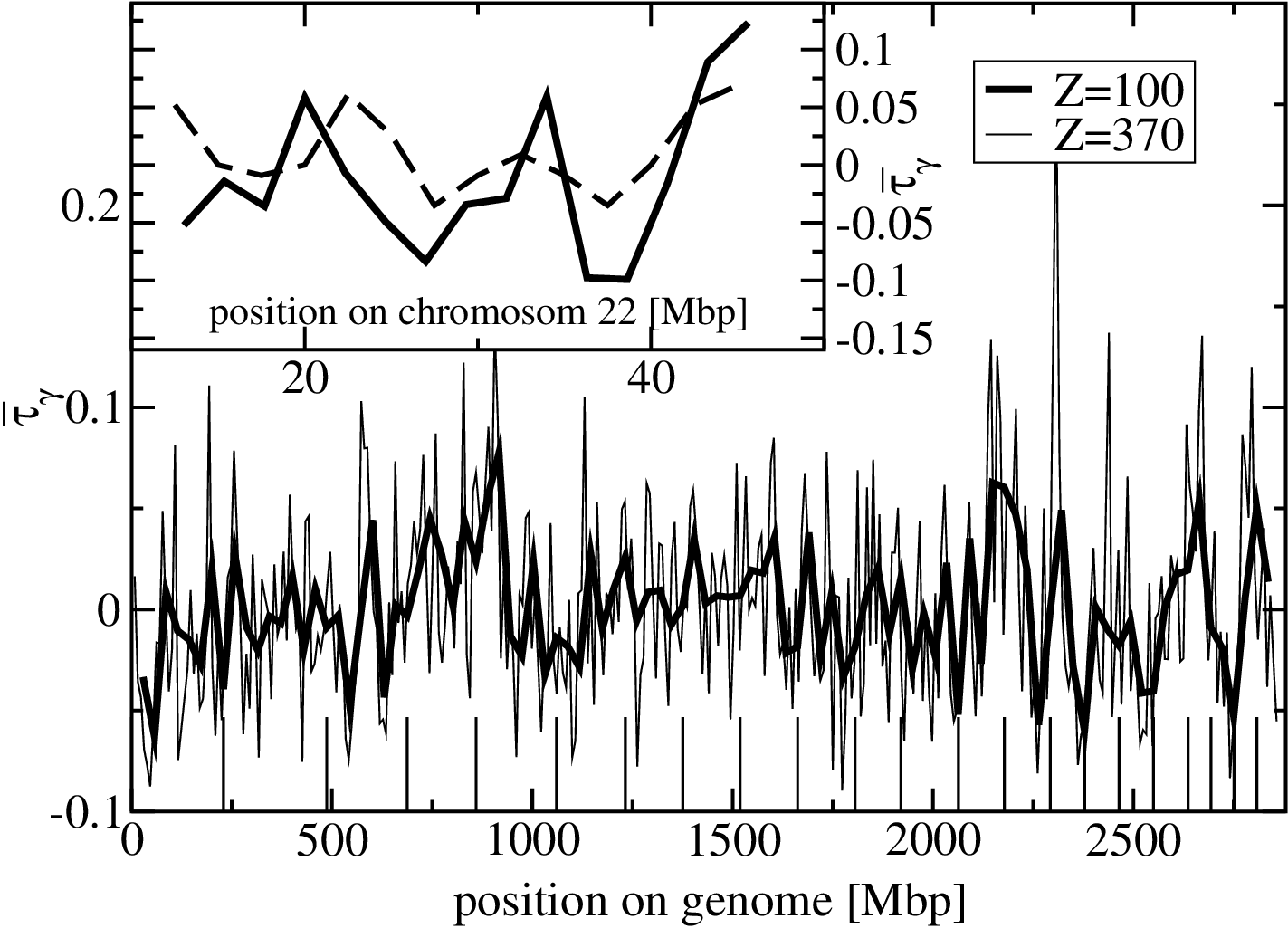}
\caption{\label{fig1} 
  Local substitution rate of the human genome for differently sized
  partitions. 
Vertical lines at the bottom
  divide the genome into the 22 chromosomes. The inset shows the our results for chromosome 22 (solid line) in comparison with the result of Ref. \cite{mouse} (dashed line).}
\end{figure}
\begin{equation}
P(k_{\alpha\gamma}|Q_{\alpha\gamma},N_{\alpha\gamma})=\frac{N_{\alpha\gamma}!\,{Q_{\alpha\gamma}}^{k_{\alpha\gamma}}(1-Q_{\alpha\gamma})^{N_{\alpha\gamma}-k_{\alpha\gamma}}}{k_{\alpha\gamma}!(N_{\alpha\gamma}-k_{\alpha\gamma})!}
\label{a2}
\end{equation}
In our case the conditional probability distributions $P(k_{\alpha\gamma}|N_{\alpha\gamma})$,
$P(Q_{\alpha\gamma}|N_{\alpha\gamma})$ are uniform for $0\leq
k_{\alpha\gamma}\leq N_{\alpha\gamma}$, $0\leq Q_{\alpha\gamma}\leq 1$.
So, within these limits, we can employ Bayes' theorem to write
$\prod_{\alpha,\gamma}P(Q_{\alpha\gamma}|k_{\alpha\gamma},N_{\alpha\gamma})=\prod_{\alpha,\gamma}P(k_{\alpha\gamma}|Q_{\alpha\gamma},N_{\alpha\gamma})$
This means, that given the sets $N_{\alpha\gamma}$, $k_{\alpha\gamma}$ we
can compute a probability distribution for $Q_{\alpha\gamma}$. 
To obtain the most probable values $\{Q_{\alpha\gamma}^{*}\}$ for the variables $\{Q_{\alpha\gamma}\}$ we have to fulfill the necessary condition
for a maximum

\begin{eqnarray}
\frac{\partial}{\partial
  Q_{\alpha\gamma}}\sum_{\alpha,\gamma}\ln P(Q_{\alpha\gamma}|k_{\alpha\gamma},N_{\alpha\gamma})\bigg|_{Q_{\alpha\gamma}
= Q^{*}_{\alpha\gamma}}&=&0
\label{a5}
\end{eqnarray}
and
$\partial_{Q_{\alpha\gamma}}^{2}\sum_{\alpha,\gamma}\ln P(Q_{\alpha\gamma}|k_{\alpha\gamma},N_{\alpha\gamma}) <0$ 
for $Q_{\alpha\gamma}=Q_{\alpha\gamma}^{*}$.
The equations~(\ref{a5}) form a set of non-linear equations.
Fortunately one can give a good estimate for values
$Q_{\alpha\gamma}^{*}$ and thus a few iterative steps using
Newton-Rapson method are sufficient to determine the Maximum Likelihood values of
the joint probability distribution, Eq.~(\ref{a2}). 
\begin{figure}
\includegraphics[width=0.45\textwidth]{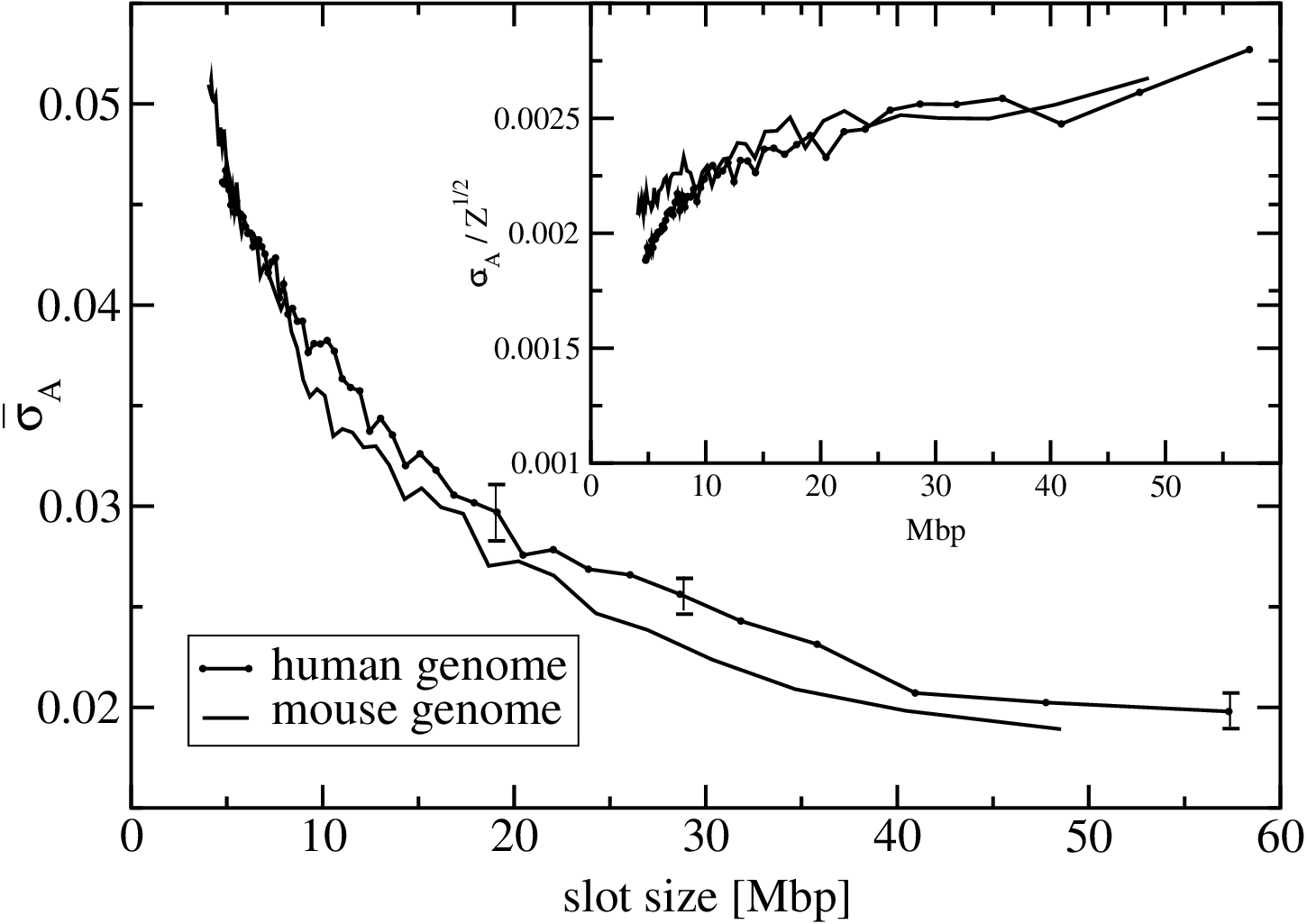}
\caption{\label{fig2} 
  Standard deviation, $\bar{\sigma}_{A}$, of the local fluctuations of
  the substitution rate versus the size of the partitions for the
  human and mouse genome. The inset shows the normalized standard
  deviations, $\bar{\sigma}_{A}/\sqrt{Z}$. The normalization,
  $\sqrt{Z}$, is chosen such that the standard deviation,
  $\bar{\sigma}_{A}$, would be constant for all partition sizes if
  $\tau_{\gamma}$ were stochastically independent. Filled squares
  represent results of a reference calculation to test the error of
  the method (see text).}
\end{figure}
To gain information from the quantities $Q_{\alpha\gamma}^{*}$ about
the quantity we are really interested in, the local substitution rate
in partition $\gamma$, denoted by $m_{\gamma}(t)$, we now introduce a
microscopic model for base substitutions.  Statistically independent
changes of a base at a given position in the genome at time $t$ can be
described by the following Master equation
\begin{equation}
\partial_{t} p_{b}(t)=-\sum_{b'}w_{bb'}(t)\,p_{b'}(t)
\label{a0}
\end{equation}
with $b,b' \in \{A,T,G,C\}$ and $p_{b}(t)$ the probability of
observing the base $b$ at that site at time $t$. The transition matrix
${\bf w}$ has the elements $w_{b b'}=-m_{\gamma}(t)\,q_{b b'}$ for
$b\neq b'$ and $w_{b b}=\sum_{b'}m_{\gamma}(t)\,q_{b b'}$ \cite{li}.
Here, the elements of the matrix ${\bf q}$, $q_{bb'}$, are the
transition probabilities that a base $b'$ mutates to a base $b$
whenever a substitution occurs. The probability that a certain base of
a repetitive element of type $\alpha$ still coincides with its
corresponding base in the consensus sequence after the time
$t_{\alpha}$ is then given by ${\bf p}(t_{\alpha})\cdot{\bf p}(0)$
(${\bf p}$ is a vector with elements $p_{b}$). The time $t_{\alpha}$
denotes the time distance from today to the moment when the
interspersed repeat of type $\alpha$ was inserted into the genome for
the first time. We emphasize that nearest neighbor effects have impact 
onto the substitution rate pattern (c.f. Ref. \cite{arndt}) but seem not to dominate the fluctuations in substitution rate on the large length scales considered \cite{mouse}. 
In the following we make the
simplifying approximation, called Jukes-Cantor approximation
\cite{li}, which amounts in taking an uniform substitution pattern,
$q_{bb'}=1/4$ for $b\neq b'$.  This is clearly a
crude approximation as, e.g., the transitions $A:T \leftrightarrow
G:C$ occur about a factor $3 - 4$ more frequently than other
substitution events \cite{arndt,ebersberger}. This approximation is
justified a posteriori, by performing Monte Carlo simulation of
synthetic data sets which include this high asymmetry in the
substitution pattern. Within our stochastical model, Eq. (\ref{a0}),
the mean divergence per base of a repetitive element of type $\alpha$
in partition $\gamma$ is given by
\begin{eqnarray}
Q^{*}_{\alpha\gamma}&=& 1-\sum_{\lambda}\exp\left[-\lambda\,\int_{0}^{t_{\alpha}}m_{\gamma}(t')dt'\right]({\bf a}_{\lambda}\cdot{\bf p}(0))^{2}\quad
\label{a1}
\end{eqnarray}
where ${\bf a}_{\lambda}$ denotes an eigenvector of the matrix ${\bf
  q}$ and $\lambda$ its eigenvalue. We define the genome-wide average
substitution rate by $m(t)=1/Z\sum_{\gamma=1}^{Z}m_{\gamma}(t)$ and
the spatial deviations from this rate as
$\tau_{\gamma}(t)=m_{\gamma}(t)-m(t)$. We further introduce the
time-averaged mean substitution rate by ${\bar
  m}_{\alpha}=1/t_{\alpha}\int_{0}^{t_{\alpha}}m(t')dt'$ and the
corresponding deviations as ${\bar\tau}_{\alpha\gamma}=1/t_{\alpha}\int_{0}^{t_{\alpha}}\tau_{\gamma}(t')dt'$.
With these abbreviations, the argument in the exponential
of Eq.~(\ref{a1}) reads ${\bar m}_{\alpha}t_{\alpha}+{\bar
  \tau}_{\alpha\gamma}t_{\alpha}=\int_{0}^{t_{\alpha}}m_{\gamma}(t')dt'$.
It is clear that we can obtain only the {\em time-averaged} quantities
$\{{\bar m}_{\alpha},\,{\bar \tau}_{\alpha\gamma}\}$ from the the knowledge of $Q_{\alpha\gamma}^{*}$.

\begin{figure}
\includegraphics[width=0.43\textwidth]{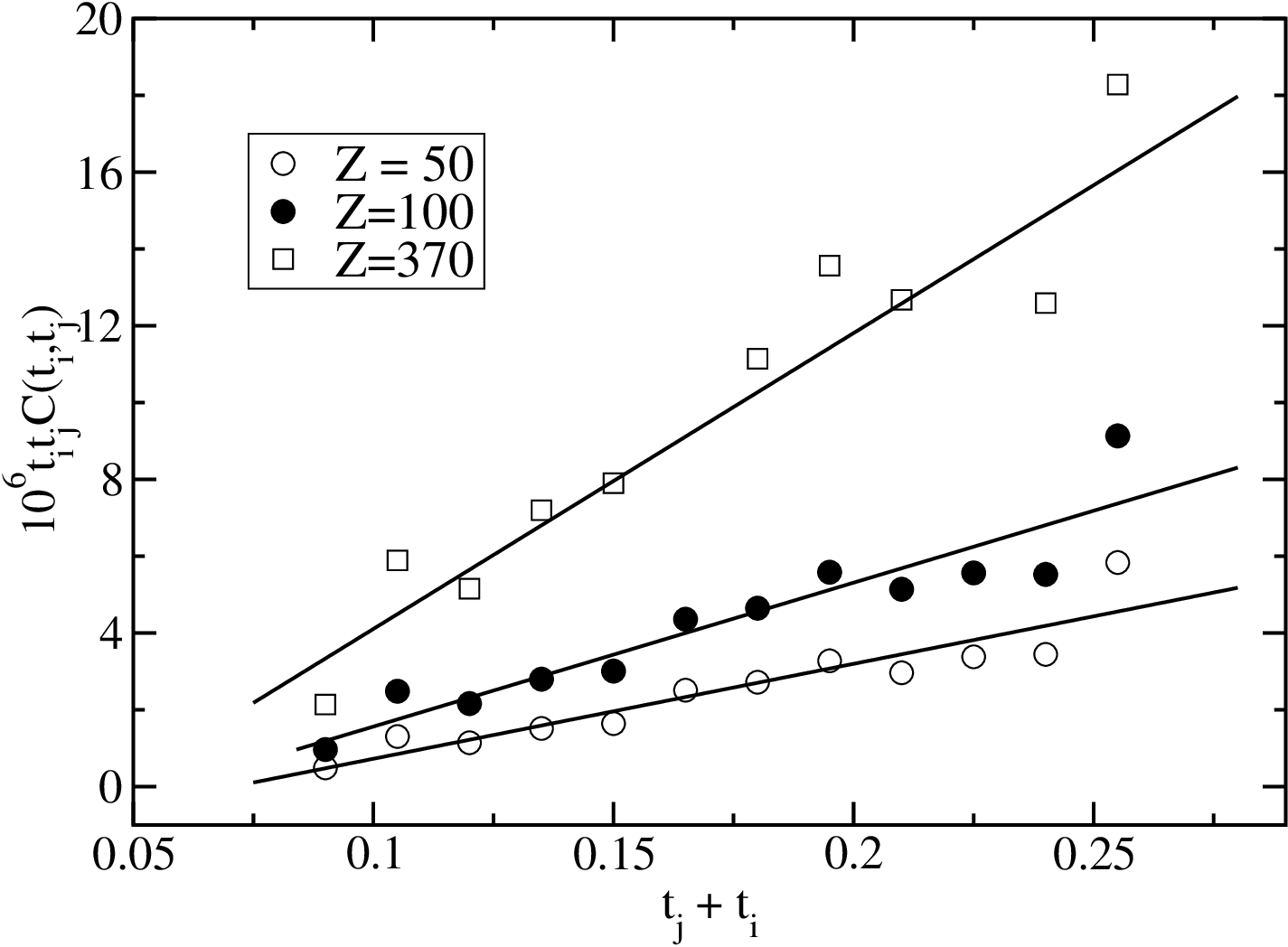}
\caption{\label{fig3} 
  The time averaged correlations function
  $t_{i}t_{j}C(t_{i},t_{j})$ versus
  $t_{i}+t_{j}$ using 8 equally sized time windows. We averaged over all values which belong to the same $t_{i}+t_{j}$. The lines are given from the first two terms of Eq.~(\ref{a8}) with the unknown variables obtained from a least mean-square fit.}
\end{figure}

The data set for interspersed repeats we use in our analysis is taken
from the UCSC Genome Bioinformatics Site \cite{ucsc}. This data set was
created using RepeatMarker together with the consecutive sequences
from the RepBase database \cite{jurka}.  The repetitive elements can
be divided into lineage-specific repeats (defined as those introduced
by transposition after the divergence of human and mouse) and
ancestral repeats (defined as those already present in the common
ancestor). In following we analyse ancient repeats, to calculate the
time averaged spatial fluctuations in substitution rate since the time
of divergence of these two species. By taking ancient repeats from a
sufficiently narrow time window we can make the approximations ${\bar
  \tau}_{\gamma}\approx {\bar \tau}_{\gamma\alpha}$, $m=m(t)$ and
thus ${\bar m}\approx {\bar m}_{\alpha}$. We set the time-scale by defining the average genome-wide substitution rate of human as ${\bar m}_{H}=1$, resulting in an average substitution rate for mouse given by ${\bar m}_{M}=2.05{\bar m}_{H}$ for ancient repeats (c.f. Ref. \cite{mouse}). As
start values for the Newton iteration scheme we take ${\bar
  \tau}_{\gamma}=0$ and $t_{\alpha}
=-1/(4{\bar m})\ln[1-(4/3)\sum_{\gamma=0}^{Z}k_{\alpha\gamma}/\sum_{\gamma=0}^{Z}N_{\alpha\gamma}]$.
Thus $Q^{*}_{\alpha\gamma}$ depends on the $Z+M$ parameters,
$\{t_{\alpha}\}$ and $\{{\bar \tau}_{\gamma}\}$ which can be
determined to high accuracy from Eq. (\ref{a5}) as shown in Fig.
(\ref{fig1}) for the human lineage using two differently sized
partitions. The standard deviations of these fluctuations, ${\bar
  \sigma}_{A}=(\sum_{\gamma=1}^{Z}{\bar \tau}_{\gamma}^{2})^{1/2}$, is
shown in Fig.~(\ref{fig2}).  The local substitution rates show
significant correlation with neighboring partitions only on length
scales smaller than $5\cdot 10^{6}$ base pairs. On larger length
scales the variations in substitution rates result mostly from
coarse-graining of statistical independent partitions of smaller size
as can be shown by rescaling the standard deviation, ${\bar
  \sigma}_{A}$, by $Z^{-1/2}$ (c.f. inset of Fig.~(\ref{fig2}). If the fluctuations were
statistically independent for the highest spatial resolution,
$Z=Z_{max}$, then
$Z^{-1/2}{\bar \sigma}_{A}$ would be constant for all $Z<Z_{max}$
and this seems to be the case for partitions of size $>5$ Mbp's ($Z<500$). By
comparing the variations in substitution rate between mouse and human
we find identical standard deviations, ${\bar \sigma}_{A}$, for both
species, Fig.~(\ref{fig2}).  This is a very surprising result as the
mean substitution rate differs by a factor two between both species
and the genome of mouse is about $14\%$ smaller than that of human.
But this agreement in the absolute magnitude could be accidental.  We
checked the bias due to the choice of repeat types by building
randomly subsets of interspersed repeats which consist just of half
the total number of repeat types used and repeating the calculations.
We also investigated the statistical errors of our genome data set
$\{k_{\alpha\gamma}\}$ by randomly creating such a data set by
Eq.~(\ref{a2}), using the readily determined values $\{t_{\alpha}\}$
and $\{{\bar \tau}_{\gamma}\}$ from the true genome data and
accounting for the large asymmetry in the substitution pattern. The
standard deviation of the combined error in determining the values
${\bar \sigma}_{A}$ is shown in Fig.~(\ref{fig2}) by the error bars for four different partitions. This validates our method and
demonstrates that the stochastic model, Eq.~(\ref{a2}), is
appropriate.\\
\indent So far, we have computed ${\bar \tau}_{\gamma}$ for a specific
time-window for the class
of youngest ancient repeats. Including also the
lineage-specific repeats, we can repeat our optimization procedure
but now with time windows including all repetitive sequences
but grouped in eight different equally distant age classes. We then approximate $\bar{\tau}_{\gamma}^{(i)}=
\int_{0}^{t_{i}}{\tau}_{\gamma}(t')\approx {\bar \tau}_{\alpha\gamma}$ for all interspersed repeats, $\alpha$, which belong to time window $i$, whose mean time distance from today is given by $t_{i}$.  We recall that
${\bar\tau}_{\gamma}^{(i)}$ is a time-averaged quantity, so it does not reflect the strength of fluctuations
of the substitution rate as found today in the human and mouse
genomes. The reason for ${\bar\tau}_{\gamma}^{(i)}$ being different
from $\tau_{\gamma}(t_{i})$ is that by
the reorganizations within the chromosomes, i.e. by insertions and
deletions of sequences, the local substitution rate can change in time. It is then clear that the
substitution rate fluctuations as shown in Fig.~(\ref{fig1}) are significantly
smaller in amplitude than the true (i.e. actual) variations in
substitution rate, $\tau_{\gamma}(t)$. To give a good estimate of the
magnitude of the true fluctuations we have to include the effect of
genome reorganizations in our model.  Within a certain large partition
$\gamma$ of size $>10$Mbp the number of repeats belonging to the same time window is sufficiently large and substitution rate across this partition
is the result of coarse graining over almost independent fluctuations
on smaller length scale. Therefore we can employ the central limit
theorem to predict that distribution of $\tau_{\gamma}(t)$ for $\gamma$ will be
close to a Gaussian distribution. This gets supported by our analysis
of ancient repeats, c.f. Fig.~(\ref{fig1}). On these length scales we
can also expect that the underlying process which changes the local
substitutions rate is Markovian as the correlation length between
different local reorganizations in the genome can be assumed to be much smaller than
partition size.
Then, assuming this process to be quasi-stationary, one is tempted to write for the actual variation of the local mutation rate, $\tau_{\gamma}(t)$, within a continuous time model
\begin{equation}
\partial_{t}\tau_{\gamma}(t)=a(t)\tau_{\gamma}(t)+\eta_{\gamma}(t)
\label{a7}
\end{equation}
This is because by Doob's theorem it is essentially the only process
satisfying the conditions stated above. Here, $a(t)$ is slowly
varying, reflecting changes in the average mutation rate and
$\eta_{\gamma}(t)$ is chosen to be white noise with zero mean. The
auto-correlation function of the process, Eq. (\ref{a7}), is
exponentially decaying. Thus, we obtain for the auto-correlation
function of the time-averaged local substitution rate, $C(t_{i},t_{j})=\frac{1}{Z}\sum_{\gamma=0}^{Z}{\bar \tau}^{(i)}_{\gamma}{\bar \tau}^{(j)}_{\gamma}$,
\begin{eqnarray}
C(t_{i},t_{j})&=&\frac{1}{\,t_{i}t_{j}}\int_{0}^{t_{i}}\int_{0}^{t_{j}}\sigma^{2}\exp[-a|t-t'|]\,dt\,dt'\label{a8}\\
&=&\frac{\sigma^{2}}{a\,t_{i}t_{j}}\bigg[t_{i}+t_{j}-|t_{i}-t_{j}|\nonumber\\
&&\quad+\frac{1}{a}\Big(e^{-\alpha t_{i}}+e^{-\alpha t_{j}}-e^{-\alpha |t_{i}-t_{j}|}-1 \Big)\bigg]\nonumber
\end{eqnarray}
with $\sigma^{2}=1/Z\sum_{\gamma=0}^{Z}\tau_{\gamma}(t)\tau_{\gamma}(t)$ the variance of the fluctuations of
actual substitution rate and $a=a(t)$. Taking the fit parameters,
$\sigma$ and $a$, time-independent is clearly an approximation but might be not a bad
one for the human species (c.f. Ref. \cite{mouse,human}).
Fig.~(\ref{fig3}) shows the correlation function, $tt'C(t,t')$, for
the human lineage. 
The free parameters $\sigma=\sigma(Z)$ and $a$ in Eq.~(\ref{a8}) are
obtained by a least mean square fit from the data. For the partition sizes resulting from
$Z=\{50,100,370\}$ we obtain the values $a=55\pm 15$ and by assuming $a=55$ we obtain from a second fit $\sigma=\{0.033\}[0.033,0.035]$ $(Z=50)$; $\sigma=\{0.043\}[0.040,0.040]$ $(Z=100)$; $\sigma=\{0.067\}[0.067,0.071]$ $(Z=370)$. The values in the brackets are obtained from the values for $\sigma$ of the two other partition sizes multiplied with the scaling factor of the two ${\bar \sigma}_{A}$'s from the corresponding partitions sizes as given in Fig.~(\ref{fig2}) e.g. $\sigma(Z_{1})={\bar \sigma}_{A}(Z_{1})\sigma(Z_{2})/{\bar \sigma}_{A}(Z_{2})$. Thus the standard deviation for the fluctuations in substitution rate is about a factor $1.7$ larger as found in the analysis using ancient repeats (c.f. Fig.~(\ref{fig2})). The time when the
correlations have decayed to $e^{-1}$ its maximum value for these
partitions is about $1/3$ of the time since
mouse and human have diverged. 
This in turn gives us an impression on
which times-scales genome reorganizations alter local substitution rates
in the human genome. We emphasize that our method can not resolve substitution rates on arbitrary small length scales as our ``measurement devices'' (repetitive elements) get shifted by insertions and deletions in other partitions over time and report therefore a spatially averaged substitution rate.\\ 
\indent In conclusion we have shown that fluctuations in substitution rate on
large length scales arise predominantly from a coarse-graining process
of fluctuations on smaller length scales.  Significant correlations
with neighboring partitions are found for the human and mouse genomes only on length scales smaller
than $5\cdot 10^{6}$ base pairs (c.f. inset Fig.~(\ref{fig2})). Moreover, both species show remarkable similarity in the the standard deviation of the fluctuations in substitution rate  on all length
scales considered.  Distinguishing between time-averaged rates and
actual substitution rates found on today's genome, we have furthermore
been able to show that the latter are significantly larger as
the time-averaged ones.  Clearly, these are the fluctuations
one has consider when trying to explain the large amount of highly
conserved sequences between the human and the mouse
genome\cite{mouse2}.

\begin{acknowledgments}
This research has been supported by Marie Curie Followship of the European Community Programme Improving the Human Research Potential and the Socio - Economic Knowledge Base under contract number HPMD-CT2001-107 in collaboration with INFM.
\end{acknowledgments}

%
\bibliography{repeats.bib}

\end{document}